\documentstyle[12pt]{article}
\newcommand{\lsim}{\raisebox{0.3mm}{\em $\, <$}
\hspace{-3.3mm} \raisebox{-1.8mm}{\em $\sim \,$}}
\begin{document}
\rightline{TMUP-HEL-9603}
\rightline{February 1996}
\baselineskip=19pt
\vskip 0.7in
\begin{center}
{\large{\bf THREE FLAVOR NEUTRINO OSCILLATION ANALYSIS OF}}
{\large{\bf THE KAMIOKANDE MULTI-GEV ATMOSPHERIC NEUTRINO DATA}}
\end{center}
\vskip 0.4in
\begin{center}
Osamu Yasuda\footnote{Email: yasuda@phys.metro-u.ac.jp}
\vskip 0.2in
{\it Department of Physics, Tokyo Metropolitan University}

{\it 1-1 Minami-Osawa Hachioji, Tokyo 192-03, Japan}
\end{center}

\vskip .7in
\centerline{ {\bf Abstract} }

Using the published Kamiokande data of the multi-GeV atmospheric neutrinos,
we have searched the optimum set of the neutrino oscillation parameters among 
three flavors.  It is found that
$\chi^2$ is minimized for ($\Delta m_{21}^2$,
$\Delta m_{31}^2$) = (3.8$\times 10^{-2}~$eV$^2$,
1.4$\times 10^{-2}~$eV$^2$), ($\theta_{12}$,
$\theta_{13}$,
$\theta_{23}$) = ($19^\circ$,$43^\circ$,$41^\circ$)
with $\chi_{\rm min}$ = 3.2 (42\%CL).
The sets of parameters ($\Delta m_{21}^2$,
$\Delta m_{31}^2$) = (${\cal O}$($10^{-11}$eV$^2$) or
${\cal O}$($10^{-5}$eV$^2$),
${\cal O}$($10^{-2}$eV$^2$)) which are suggested by the two flavor
analysis fall within 0.7$\sigma$.

\newpage

There has been much interest in atmospheric neutrinos
\cite{kamioka1} \cite{kamioka2} \cite{imb} \cite{frejus} \cite{nusex}
\cite{soudan2}, 
which might give us an evidence for neutrino oscillations.  While
NUSEX \cite{nusex} and Frejus \cite{frejus} have reported consistency
between the data and the predictions on atmospheric neutrino flux
\cite{flux} \cite{hkkm},
Kamiokande, IMB and Soudan-2 have reported discrepancy.
In particular, the Kamiokande group claimed that their multi-GeV data
suggests the mass squared difference of neutrino is of
order 10$^{-2}$eV$^2$ \cite{kamioka2}.

People have studied neutrino oscillations among three
flavors \cite{kp}, and it has been shown recently \cite{3nu}
that the mass squared differences and the mixing
angles have strong constraints from various experiments.  The analysis
of the multi-GeV atmospheric neutrino data by the Kamiokande group
\cite{kamioka2} was based on the framework of neutrino oscillation
between two flavors\footnote{A couple of works \cite{fl} \cite{s}
have discussed Kamiokande's analysis from the viewpoints
which are different from ours.},
and it is important to see what happens if we
analyze the data in the three flavor framework.  In this paper we will
analyze the published multi-GeV data \cite{kamioka2} of the Kamiokande
atmospheric neutrino experiment, taking into account mixings among
three flavor neutrinos.  Unlike other works in Ref. \cite{3nu},
we will take the matter effect \cite {msw} into consideration,
and evaluate the number of events by summing over the energy and
the zenith angle of neutrinos, to reproduce the original analysis
by the Kamiokande group as much as possible.
Throughout this paper we will restrict our
discussions only to the multi-GeV data by the Kamiokande group, not only
because the Monte Carlo result for the neutrino energy spectrum
is available only in Ref. \cite{kamioka2}, but also because
this is the only data which gives both the upper and the lower bound
on the mass squared difference of neutrinos.

We start with the Dirac equation for three flavors of neutrinos
with mass in matter \cite{msw}:
\begin{eqnarray}
i\frac{d}{dx} \Psi (x) =
\left[ U {\rm diag}\left( E_1,E_2,E_3 \right) U^{-1} + 
{\rm diag}\left( A(x),0,0 \right)
\right] \Psi (x)
\label{eqn:dirac}
\end{eqnarray}

Here $E_j=\sqrt{p^2+m_j^2}$ is the energy of the neutrino,
$\Psi(x) \equiv (\nu_e(x), \nu_\mu(x), \nu_\tau(x))^T$ is
the wave function of the neutrinos in the flavor basis,
$A(x) \equiv \sqrt{2} G_F N_e(x)$~stands for the effect
due to the charged current interactions between $\nu_e$
and electrons in matter \cite{msw}.
\begin{eqnarray}
U&\equiv&\left(
\begin{array}{ccc}
U_{e1} & U_{e2} &  U_{e3}\\
U_{\mu 1} & U_{\mu 2} & U_{\mu 3} \\
U_{\tau 1} & U_{\tau 2} & U_{\tau 3}
\end{array}\right)\nonumber\\
&\equiv&\left(
\begin{array}{ccc}
c_{12}c_{13} & s_{12}c_{13} &   s_{13}\\
-s_{12}c_{23}-c_{12}s_{23}s_{13} & c_{12}c_{23}-s_{12}s_{23}s_{13}
& s_{23}c_{13}\\
s_{12}s_{23}-c_{12}c_{23}s_{13} & -c_{12}s_{23}-s_{12}c_{23}s_{13}
& c_{23}c_{13}
\end{array}
\right)
\end{eqnarray}
with $c_{ij}\equiv\cos\theta_{ij},~s_{ij}\equiv\sin\theta_{ij}$
is the orthogonal mixing matrix of neutrinos, and we will not
discuss the CP violating phase of the mixing matrix here for simplicity
\footnote{Even if we include the CP violating phase $\delta$
of the mixing matrix, the effect of $\delta$ always appears
in the combination of $s_{13}e^\delta$.  $s_{13}$ has
to be small because of the constraints from the reactor experiments
and from the solar neutrino observations, as we will discuss
below.}.

The number of the expected charged leptons $\ell_\alpha$~
( $\ell_\alpha$ = $\mu$ or
$\tau$ ) with energy $q$ from a scattering $\nu_\alpha e 
\rightarrow \nu_e \ell_\alpha$
is given by
\begin{eqnarray}
\displaystyle
N(\ell_\alpha)
&=& n_T\sum_{\beta=e,\mu}
\int_0^\infty dE
\int_0^\pi d\Theta
\int^{q_{\rm max}}_0 dq\nonumber\\
&{ }&\epsilon (q)
F_\beta (E,\Theta) { d\sigma_\alpha (E,q) \over dq }
P(\nu_\beta\rightarrow\nu_\alpha; E, \Theta)
\quad (\alpha=e,\mu)
\end{eqnarray}
Here $F_\beta (E, \Theta)$ is the flux of atmospheric neutrino
$\nu_\beta$ with energy $E$ from the zenith angle $\Theta$,
$n_T$ is
the effective number of target nucleons, $\epsilon (q)$ is
the detection
efficiency function for charged leptons $\ell_\alpha$,
$d\sigma_\alpha(E,q)/dq$ is the differential cross section
of the interaction $\nu_\alpha e^- \rightarrow \nu_e \ell_\alpha^-$
($\alpha$ = $e$ or $\mu$),
$P(\nu_\beta\rightarrow\nu_\alpha; E, \Theta)\equiv|\langle\nu_\beta
(0)|\nu_\alpha(L)\rangle|^2$ is the probability
of $\nu_\beta\rightarrow\nu_\alpha$ transitions with energy $E$ after
traveling a distance
\begin{eqnarray}
\displaystyle L=\sqrt{(R+h)^2-R^2\sin^2\Theta}-R\cos\Theta,
\end{eqnarray}
where $R$ is the radius of the Earth, $h\sim$15Km is the altitude
at which atmospheric neutrinos are produced.

To reproduce the analysis of the multi-GeV data by the Kamiokande group,
one needs the quantity
\begin{eqnarray}
f_{\beta\alpha} (E,\Theta)\equiv
 n_T\int^{q_{\rm max}}_0 dq~\epsilon (q)
F_\beta (E,\Theta) { d\sigma_\alpha (E,q) \over dq }
\quad (\alpha,\beta=e,\mu)
\label{eqn:flux0}
\end{eqnarray}
for each $E$ and $\Theta$, which is not given in \cite{kamioka2}.
However, the quantity 
\begin{eqnarray}
g_{\beta\alpha} (E) \equiv \int_0^\pi d\Theta~f_{\beta\alpha} (E,\Theta),
\end{eqnarray}
which is obtained by integrating (\ref{eqn:flux0})
over $\Theta$, is given in the Fig.2 (d)--(f) in Ref. \cite{kamioka2}.
The zenith angle dependence $n_\beta (E, \Theta)$ of the atmospheric
neutrino flux for various neutrino energy $E$ has been given
in Ref. \cite{hkkm} in detail.  Here we multiply the quantity
$g_{\beta\alpha} (E)$
by the zenith angle dependence in Ref. \cite{hkkm} with suitable
normalization, and adopt the quantity
\begin{eqnarray}
\widetilde f_{\beta\alpha} (E,\Theta)\equiv
{g_{\beta\alpha} (E) n_\beta (E, \Theta) \over
\int_0^\pi d\Theta ~n_\beta (E,\Theta)}
\quad (\alpha,\beta=e,\mu)
\label{eqn:flux}
\end{eqnarray}
instead of the original quantity $f_{\beta\alpha}
(E,\Theta)$ used in \cite{kamioka2}.  (\ref{eqn:flux}) is the
important assumption of the present analysis.  (\ref{eqn:flux}) is not
exactly the same as $f_{\beta\alpha}(E,\Theta)$ in the original analysis
\cite{kamioka2}, but this is almost the best which can
be done with the published data in \cite{kamioka2}.

We have solved (\ref{eqn:dirac}) numerically for each $E$
(10$^{-1/20}$ GeV $\le$ $E$ $\le$ 10$^2$ GeV) and evaluated
the number of events for a given range of the zenith angle
$\Theta_j\equiv\cos^{-1}({2j-7 \over 5}) < \Theta <
\Theta_{j+1}\equiv\cos^{-1}({2j-5 \over 5})$ ($1 \le j \le 5$)
\begin{eqnarray}
X^\mu_j &\equiv& N(\mu,\Theta_j<\Theta<\Theta_{j+1})\nonumber\\
&\equiv&(1+\alpha)(1+{\beta \over 2})
\int_{\Theta_j}^{\Theta_{j+1}} d\Theta \int dE\nonumber\\
&{ }&\left[ \widetilde f_{\mu\mu} (E,\Theta)
P(\nu_\mu\rightarrow\nu_\mu;E,\Theta)
+\widetilde f_{e\mu} (E,\Theta) P(\nu_e\rightarrow\nu_\mu;E,\Theta)
\right]\nonumber\\
X^e_j &\equiv& N(e,\Theta_j<\Theta<\Theta_{j+1})\nonumber\\
&\equiv&(1+\alpha)(1-{\beta \over 2})
\int_{\Theta_j}^{\Theta_{j+1}} d\Theta \int dE\nonumber\\
&{ }&\left[ \widetilde f_{\mu e} (E,\Theta) P(\nu_\mu\rightarrow\nu_e;E,\Theta)
+\widetilde f_{ee} (E,\Theta) P(\nu_e\rightarrow\nu_e;E,\Theta)
\right].
\end{eqnarray}
Several groups \cite{flux} \cite{hkkm} have given predictions on the
flux of atmospheric neutrinos but they differ from one another in the
magnitudes, and the Kamiokande group assumed that the errors of the
overall normalization $1+\alpha$ and the relative normalization $1+\beta/2$
are $\sigma_\alpha$=30\% and $\sigma_\beta$=12\%, respectively.  Here
we regard these factors $\alpha$ and $\beta$ as free parameters of the
theory, and adopt the following $\chi^2$ \cite{bc}:
\begin{eqnarray}
\chi^2=2\sum_{\alpha=e,\mu}\sum_{j=1}^5
\left( X_j^\alpha-N_j^\alpha-N_j^\alpha
\ln{X_j^\alpha \over N_j^\alpha} \right),
\label{eqn:chi}
\end{eqnarray}
where $N_j^\alpha~(\alpha=e,\mu;1\le j \le 5)$ is the data
for each zenith angle $\Theta_j<\Theta<\Theta_{j+1}$.
The theoretical prediction $X_j^\alpha~(\alpha=e,\mu;1\le j \le 5)$
depends on seven free parameters ($\Delta m_{21}^2$,$\Delta m_{31}^2$;
$\theta_{12}$,$\theta_{13}$,$\theta_{23}$;~$\alpha$,$\beta$),
where $\Delta m_{ij}\equiv m_i^2-m_j^2$, so (\ref{eqn:chi})
is expected to obey a $\chi^2$ distribution with 10$-$7=3 degrees
of freedom.  The number of degrees of freedom in the present
analysis is smaller than the original one by the Kamiokande
group (5$\times$8+5$-$2=83). 

We could not reproduce exactly the zenith angle distributions in Fig. 
3 in Ref. \cite{kamioka2}.  In particular our prediction for e-like
events near $\cos\Theta\sim -1$ has a larger difference with the data
than Kamiokande's does, and this difference seems to be important to
discuss the magnitude of $\chi^2$ later\footnote{If we try to fit the
data with only two parameters $\alpha$ and $\beta$ as in Ref.
\cite{s}, then the minimum value of $\chi^2$ is 21, which suggests
that the $\Theta$ independent solution is excluded at the
99\% confidence level in our analysis.}.
Presumably this discrepancy
arises not only because the data that we are using is different from
the original one in Ref. \cite{kamioka2}, but also because the
Kamiokande group has
taken into account the smearing effect on the resolution of the angle
($15^\circ \sim 20^\circ$) and the effects of backgrounds
\cite{kamioka2} \cite{kajita}.  Throughout this paper we discuss
the goodness of fit and the confidence level of set of the
parameters etc. based on our calculation with (\ref{eqn:flux}).

The value of $\chi^2$ is affected to some extent by the presence
of matter, and it is necessary to take into consideration
the contribution of the second term in (\ref{eqn:dirac}).
Evaluation of $\chi^2$ requires a lot of CPU time of a computer
since one has to solve (\ref{eqn:dirac}) numerically for each
$E$ and $\Theta$ and plug it into (\ref{eqn:chi}).  We have meshed
each parameter region into ten points ($\Delta m_{ij}^2
=10^{-5+\ell/2},~\theta_{ij}=\ell\pi / 20~(0\le \ell \le 10)$)
and the evaluated the value of $\chi^2$.  Furthermore, using the
gradient-search method described in Ref. \cite{br}, we have
found that $\chi^2$ has the minimum value for
\begin{eqnarray}
&{ }&(\Delta m_{21}^2,\Delta m_{31}^2)=
(3.8\times 10^{-2}{\rm eV}^2,1.4\times 10^{-2}{\rm eV}^2)\nonumber\\
&{ }&(\theta_{12},\theta_{13},\theta_{23})=
(19^\circ,43^\circ,41^\circ)\nonumber\\
&{ }&(\alpha,\beta)=(2.8\times10^{-1},-5.0\times10^{-2})
\end{eqnarray}
with $\chi^2_{\rm min}=3.2.$ Note that the deviation of the two
normalization factors $1+\alpha$ and $1+\beta/2$ from unity is within the
errors $\sigma_\alpha$=30\% $\sigma_\beta$=12\% assumed in Ref.
\cite{kamioka2}.  We have also calculated the value of the
modified chi square $\widetilde \chi^2\equiv \chi^2
+\alpha^2/\sigma^2_\alpha +\beta^2/\sigma^2_\beta$ with the weight for
the errors $\sigma_\alpha$=30\% $\sigma_\beta$=12\%, and we have found
that the conclusions in the following discussions do not change very
much with $\widetilde \chi^2$ instead of $\chi^2$.

The zenith angle distributions of the e-like events, the $\mu$-like
events and the double ratio $R\equiv(\mu/e)_{\rm data}/(\mu/e)_{\rm MC}$
are given in Fig.1 and Fig.2 for the optimum set of parameters.

\vglue 0.5truecm
(Insert Fig.1 and Fig.2 here.)
\vglue 0.5truecm

The degrees of freedom of our analysis is 3, so the value of the
reduced chi square is 1.1, which corresponds to 42 \% confidence
level.  This suggests that our fit in the present analysis is not
particularly good, but as we mentioned earlier, this is probably due
to the fact that the data from which we start is poorer than the
original one by the Kamiokande group \cite{kamioka2}.

The region of the parameter space which is allowed at 90 \% CL is
given by $\chi^2 \le \chi^2_{\rm min}+12$ for seven free parameters.
However, because it requires a lot of CPU time of a computer
to solve (\ref{eqn:dirac}) numerically, we could not give 
sets of contours in the parameter space.  In fact,
since the value $\chi-\chi_{\rm min}$ is rather large, if we project
the allowed region onto the $\Delta m_{21}^2 - \Delta m_{31}^2$
plane, it is conceivable that we have disjointed regions on this plane,
and the calculation would be extremely tedious.
So we restrict our analysis
to a special case of particular interest here.

We have evaluated $\chi^2$ for the sets of parameters, which are
suggested by the solutions for the solar neutrino problem \cite{msw}
\cite{solar}.  In order not to spoil the success of these
scenarios  \cite{msw}
\cite{solar} based on the two flavor framework, we consider only
the case in which $|U_{e3}|$ is small, since we have the formula which
relates the probability $P^{(3)}(\nu_e\rightarrow\nu_e)$ in the three
flavor analysis to $P^{(2)}(\nu_e\rightarrow\nu_e)$ in the two flavor
one \cite{3to2}:
\begin{eqnarray}
P^{(3)}(\nu_e\rightarrow\nu_e;A(x))
=(1-|U_{e3}|^2)^2 P^{(2)}
(\nu_e\rightarrow\nu_e;(1-|U_{e3}|^2)A(x))+|U_{e3}|^4.
\end{eqnarray}
We note in passing that the constraint for $|U_{e3}|$ also
comes from the reactor experiments \cite{bugey}, which
suggests $|U_{e3}|^2=s_{13}^2\lsim  10^{-1}$ or
$1-|U_{e3}|^2=c_{13}^2\lsim  10^{-1}$ for
$\Delta m_{31}^2\simeq$ a few 10$^{-2}$eV$^2$.

Irrespective of whether we consider the vacuum solution
($\Delta m_{21}^2\sim{\cal O}(10^{-11}$eV$^2$)) or
the MSW solution ($\Delta m_{21}^2\sim{\cal O}(10^{-5}$eV$^2$))
for the solar neutrino,
the mass squared difference $\Delta m_{21}^2$ is negligible compared to
the contribution of the matter effect $A(x)$ in (\ref{eqn:dirac}) and
the other mass squared difference $\Delta m_{31}^2$, which should be at least
of order 10$^{-2}$eV$^2$ to account for the zenith angle
dependence of the Kamiokande multi-GeV data.  Thus we consider the case
\begin{eqnarray}
&{ }&(\Delta m_{21}^2,\sin^22\theta_{12})
=(\Delta m^2,\sin^22\theta)_\odot\nonumber\\
&{ }&\equiv\left\{ \begin{array}{lr}
({\cal O}(10^{-11}{\rm eV}^2),{\cal O}(1)),&  ({\rm vacuum~solution})\\
({\cal O}(10^{-5}{\rm eV}^2),{\cal O}(10^{-2})),&
({\rm small~angle~MSW~solution})\\
({\cal O}(10^{-5}{\rm eV}^2),{\cal O}(1)),&
({\rm large~angle~MSW~solution})
	     \end{array} \right.\nonumber\\
&{ }& 0\le\theta_{13}\lsim 5^\circ\nonumber\\
&{ }& \Delta m_{31}^2,\theta_{23}={\rm arbitrary}
\label{eqn:param}
\end{eqnarray}
With this constraint we have found that $\chi^2$ has the minimum value
for
\begin{eqnarray}
\Delta m_{31}^2&=&3\times 10^{-2}{\rm eV}^2\nonumber\\
\theta_{13}&=&5^\circ\nonumber\\
\theta_{23}&=&40^\circ~{\rm or}~50^\circ\nonumber\\
(\alpha,\beta)&=&(3.6\times10^{-1},-3.7\times10^{-2})
\label{eqn:sol}
\end{eqnarray}
with
\begin{eqnarray}
\chi^2=9.6.
\end{eqnarray}
Deviation of each parameter in this case is
$(\chi^2-\chi^2_{\rm min})/7=0.9$,
so we conclude that this set of parameters falls within
$0.7\sigma$ for all three cases in (\ref{eqn:param}).
In fact we observe that any set of the parameters with the
constraints (\ref{eqn:sol}) falls within $0.7\sigma$
as long as $\Delta m_{21}^2\ll\Delta m_{32}^2<\Delta m_{31}^2$
is satisfied (arbitrary $\theta_{12}$ is allowed in this case).
In (\ref{eqn:sol}) the error of $\alpha$ is
a little too large compared to what the Kamiokande group
assumed, but even if we take $\alpha=3.0\times10^{-1}$
with all other parameters the same as in (\ref{eqn:sol}),
we find that the solution falls within $0.8\sigma$.
The reason that we have weaker constraints in this analysis
than in Ref. \cite{kamioka2} is because we have larger
numbers of free parameters, but this is inevitable
as long as one assumes the general mixings among three flavors
of neutrinos.

In this paper we have analyzed the multi-GeV atmospheric neutrino
data by the Kamiokande group based on the framework of
three flavor neutrino oscillations, and have shown that the
best fit is obtained for the set of parameters
$\Delta m_{21}^2\sim\Delta m_{31}^2\sim {\cal O}$ (10$^{-2}$eV$^2$).
We have also shown that the popular set of parameters
( $(\Delta m_{21}^2,\sin^2 2\theta_{12})=(\Delta m^2,\sin^2 2\theta)_\odot$,
$(\Delta m_{31}^2,\sin^2 2\theta_{13})=({\cal O}(10^{-2}$eV$^2$),
${\cal O}(1)$), $\theta_{13}\simeq 0$ ) fall within 0.7$\sigma$.
The minimum value of $\chi^2$ is 3.2 for 3 degrees of freedom,
and the fit based on the hypothesis of neutrino oscillations
is not particularly good due to that fact that we used only the
information published in Ref. \cite{kamioka2}.  We hope that
the situation will be improved much more when the SuperKamiokande
experiment starts.  If we combine the results here with other
experimental data, then we get even stronger constraints,
which will be reported somewhere \cite{my}.

\vskip 0.2in
\noindent
{\Large{\bf Acknowledgement}}
\vskip 0.1in

The author would like to thank H. Minakata for discussions and
comments on the manuscript, K.S. Babu, P.I. Krastev, C.N. Leung, A. 
Smirnov, and L. Volkova for discussions, and T. Kajita for a useful
communication.  He also would like to thank members of the Physics
Department of Yale University for their hospitality during part of
this work.  This research was supported in part by a Grant-in-Aid of
Education, Science and Culture, \#05302016, \#05640355, \#07044092.

\newpage
\noindent
{\Large{\bf Figures}}

\begin{description}
\item[Fig.1 (a),(b)] Zenith angle distributions for the e-like
and $\mu$-like multi-GeV events.
The squares with error bars are data and the histograms stand for the
predictions without neutrino oscillations (solid lines), and
with neutrino oscillations (dashed lines for the optimum
set of parameters ($\Delta m_{21}^2$,
$\Delta m_{31}^2$) = (3.8$\times 10^{-2}~$eV$^2$,
1.4$\times 10^{-2}~$eV$^2$), ($\theta_{12}$,~$\theta_{13}$,~$\theta_{23}$)
= ($19^\circ$,$43^\circ$,$41^\circ$)), respectively.
These quantities are obtained by multiplying the values in Fig.3(d)
in Ref. \cite{kamioka2} by the zenith angle dependence of the flux
in Ref. \cite{hkkm}.

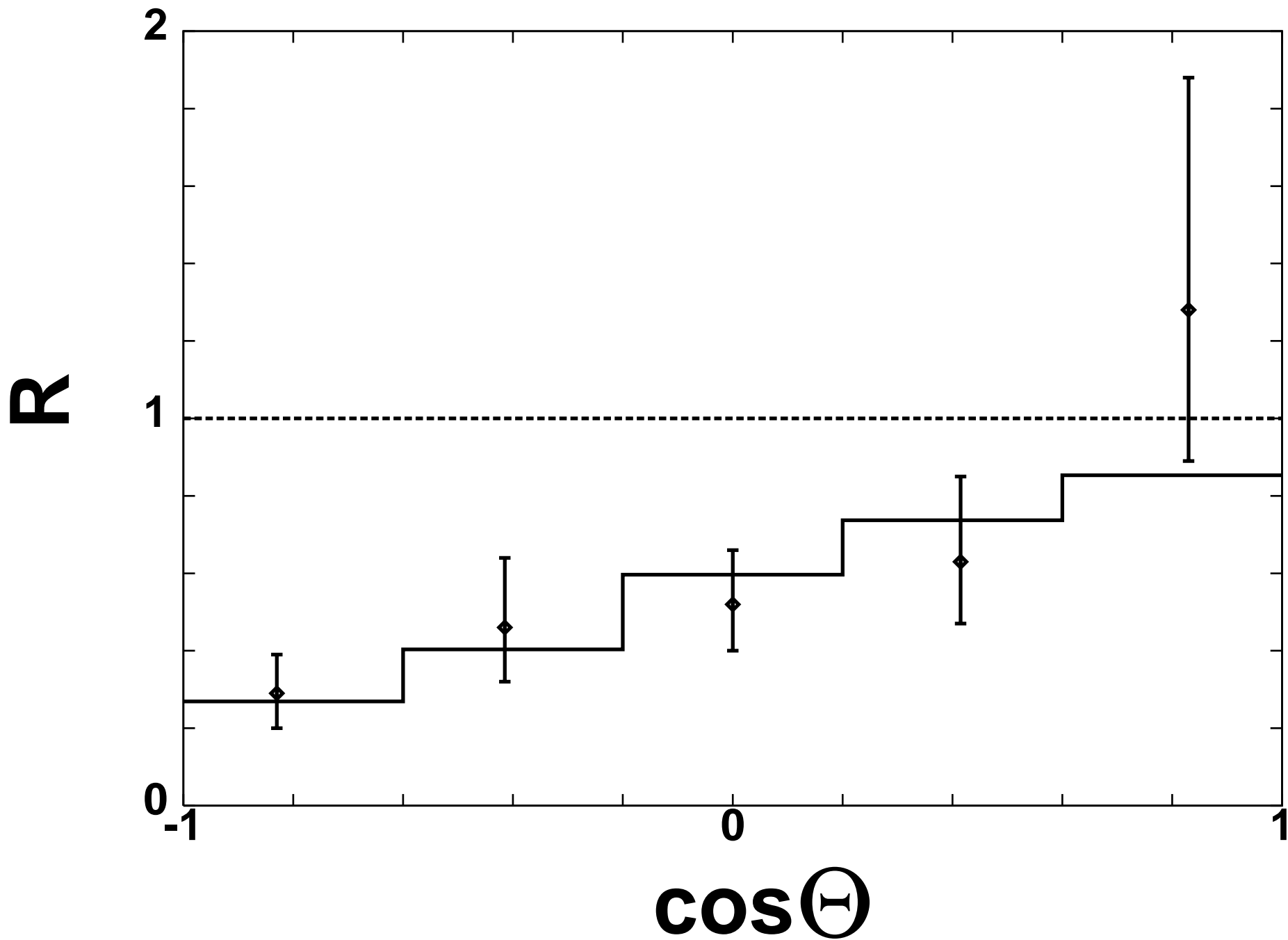
\item[Fig.2] Zenith angle distribution of the double ratio
$R\equiv(\mu/e)_{\rm data}/(\mu/e)_{\rm MC}$.  The solid lines stand for the
prediction with neutrino oscillations for the optimum set of
parameters.  All the quantities are calculated based on the same assumption
as in Fig.1.

\end{description}


\vskip 0.2in
\begin{thebibliography}{99}
\bibitem{kamioka1}Kamiokande Collaboration, K.S. Hirata et al.,  Phys. Lett. {\bf B205} (1988) 416;
{\it ibid.} {\bf B280} (1992) 146. 
\bibitem{kamioka2}
Kamiokande Collaboration, Y. Fukuda et al., Phys. Lett. {\bf B335} (1994) 237. 
\bibitem{imb}
IMB Collaboration, D. Casper et al., Phys. Rev. Lett. {\bf 66} (1989) 2561;
R. Becker-Szendy et al., Phys. Rev. {\bf D46} (1989) 3720.
\bibitem{nusex}
NUSEX Collaboration, M. Aglietta et al., Europhys. Lett. {\bf 8} (1989) 611.
\bibitem{frejus}
Frejus Collaboration, Ch. Berger et al.,  Phys. Lett. {\bf B227} (1989) 489;
{\it ibid.} {\bf B245} (1990) 305; K. Daum et al, Z. Phys. {\bf C66}
(1995) 417. 
\bibitem{soudan2}
Soudan 2 Collaboration, M. Goodman et al., Nucl. Phys. {\bf B}
(Proc. Suppl.) {\bf 38} (1995) 337.
\bibitem{hkkm}
M. Honda, T. Kajita, S. Midorikawa, and K. Kasahara,
Phys. Rev. {\bf D52} (1995) 4985.
\bibitem{flux}
L.V. Volkova, Sov. J. Nucl. Phys. {\bf 31} (1980) 784;
T.K. Gaisser, T. Stanev S.A. Bludman and H. Lee, Phys. Rev. Lett.
{\bf 51} (1983) 223;
A. Dar, Phys. Rev. Lett.
{\bf 51} (1983) 227;
K. Mitsui, Y. Minorikawa and H. Komori, Nuovo Cim. {\bf C9} (1986) 995;
E.V. Bugaev and V.A. Naumov, Sov. J. Nucl. Phys. {\bf 45} (1987) 857;
T.K. Gaisser, T. Stanev and G. Bar, Phys. Rev. {\bf D38} (1988) 85;
A.V. Butkevich, L.G. Dedenko and I.M. Zheleznykh, Sov. J. Nucl. Phys.
{\bf 50} (1989) 90;
M. Honda, K. Kasahara, K. Hidaka and S. Midorikawa, Phys. Lett. {\bf B248},
193 (1990);
H. Lee and Y. S. Koh, Nuovo Cim. {\bf B105} (1990) 883;
M. Kawasaki and S. Mizuta, Phys. Rev. {\bf D43} (1991) 2900;
P. Lipari, Astropart. Phys. {\bf 1} (1993) 195.
D.H. Perkins, Astropart. Phys. {\bf 2} (1994) 249;
V. Agrawal, T.K. Gaisser, P. Lipari and T. Stanev preprint BA-95-49
(hep-ph/9509423).
\bibitem{kp}
See T.K. Kuo and J. Pantaleone, Rev. Mod. Phys. {\bf 61} (1989) 937
and references therein.
\bibitem{3nu}
A. Acker, A.B. Balantekin, F. Loreti, Phys. Rev. {\bf D49} (1994) 328;
J. Pantaleone, Phys. Rev. {\bf D49} (1994) R2152;
G.L. Fogli, E. Lisi, D. Montanino, Phys. Rev. {\bf D49} (1994) 3626;
H. Minakata, Phys. Lett. {\bf B356} (1995) 61;
Phys. Rev. {\bf D52} (1995) 6630;
S.M. Bilenky, A. Bottino, C. Giunti, C.W. Kim,
Phys. Lett. {\bf B356} (1995) 273; preprint DFTT 2/96 (hep-ph/9602216);
S.M. Bilenky, C. Giunti, C.W. Kim, preprint DFTT-30-95
(hep-ph/9505301);
M. Narayan, M.V.N. Murthy, G. Rajasekaran, S. Uma Sankar,
preprint IMSC-95-96-001 (hep-ph/9505281);
G.L. Fogli, E. Lisi and G. Scioscia, Phys. Rev. {\bf D52} (1995) 5334;
S. Goswami, K. Kar and A. Raychaudhuri, preprint CUPP-95-3 (hep-ph/9505395).
\bibitem{fl}
G.L. Fogli, E. Lisi, Phys. Rev. {\bf D52} (1995) 2775.
\bibitem{s}
D. Saltzberg, Phys. Lett. {\bf B355} (1995) 499.
\bibitem{msw}
S. P. Mikheyev and A. Smirnov, Nuovo Cim. {\bf 9C} (1986) 17; 
L. Wolfenstein, Phys. Rev. {\bf D17} (1978) 2369.
\bibitem{bc}
S. Baker and R.D. Cousin, Nucl. Instr. and Meth., {\bf 221} (1984) 437.
\bibitem{kajita}T. Kajita, private communication.
\bibitem{br}
P.R. Bevington and D.K. Robinson,
{\sl DATA REDUCTION AND ERROR ANALYSIS FOR THE PHYSICAL SCIENCES},
2nd ed. N.Y., McGraw-Hill, 1992.
\bibitem{solar}
See, e.g., J.N. Bahcall and R.K. Ulrich,
Rev. Mod. Phys. {\bf 60} (1988) 297;
J.N. Bahcall and M.H. Pinsonneault,
Rev. Mod. Phys. {\bf 64} (1992) 885;
J.N. Bahcall, R. Davis, Jr., P. Parker, A. Smirnov, R. Ulrich
eds., {\sl SOLAR NEUTRINOS: the first thirty years}
Reading, Mass., Addison-Wesley, 1994 and references therein.
\bibitem{3to2}
C.-S. Lim, Proc. of the
BNL Neutrino Workshop on Opportunities for Neutrino Physics at BNL,
Upton, N.Y., February 5-7, 1987, ed. by M. J. Murtagh, p111;
A. Yu. Smirnov, Proc. of the Int Symposium on
Neutrino Astrophysics, Takayama/Kamioka 19 - 22 October 1992,
ed. by Y. Suzuki and K. Nakamura, p.105.
\bibitem{bugey}
B. Ackar et al., Nucl. Phys. {\bf B434}, (1995) 503.
\bibitem{my}
O. Yasuda and H. Minakata, preprint TMUP-HEL-9604.
\end{thebibliography}
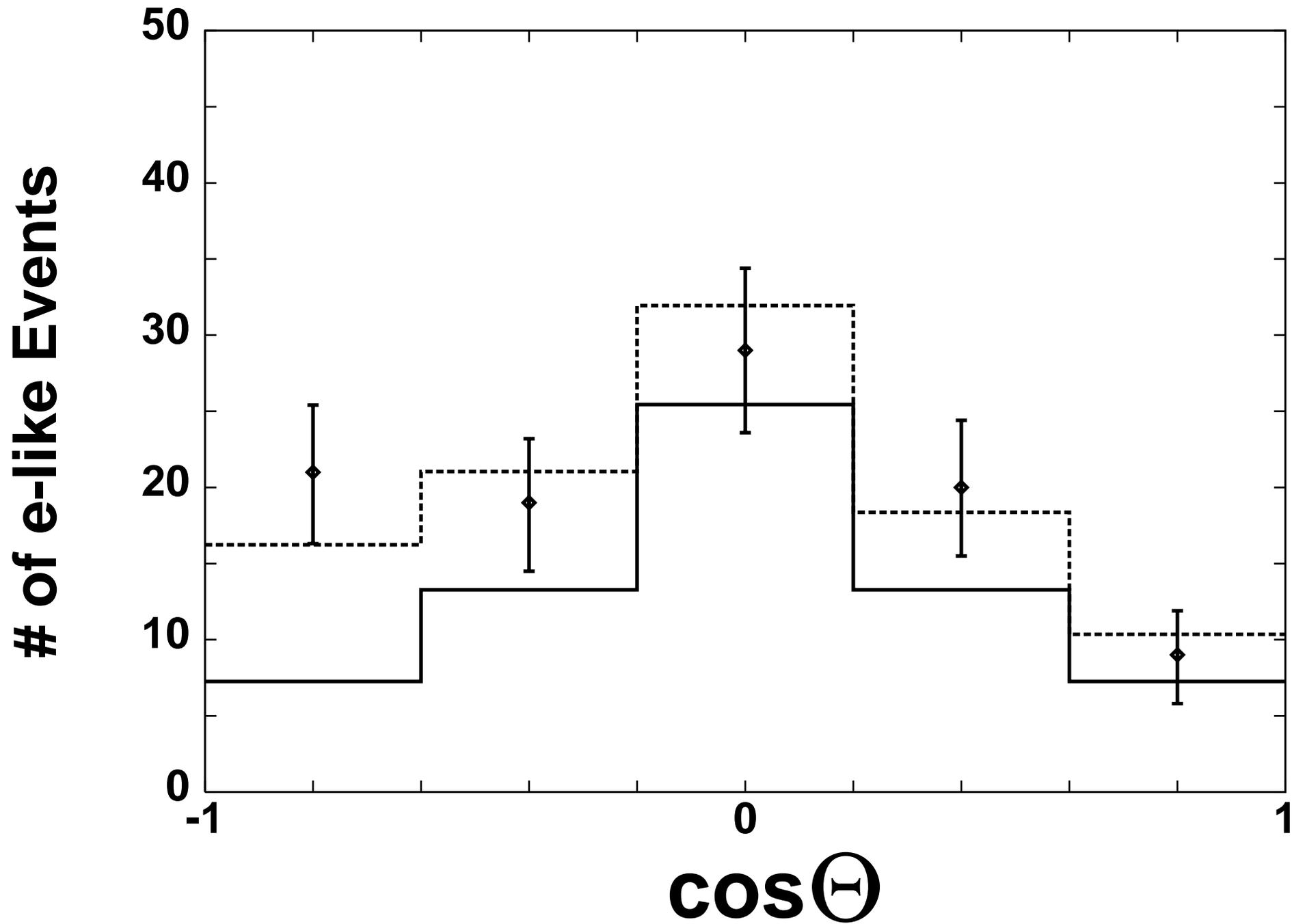
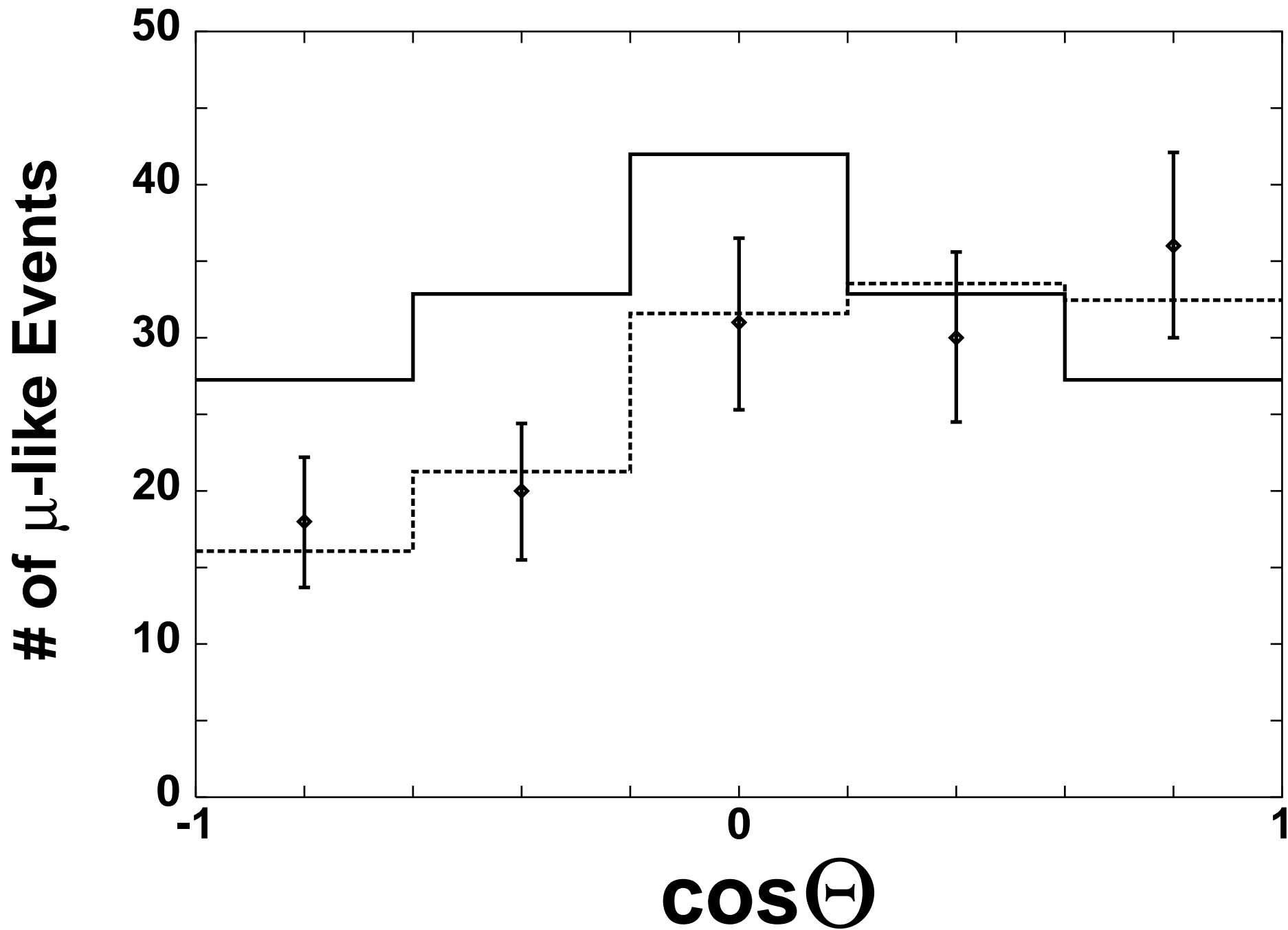
\end{document}